\documentclass[aps,pra,twocolumn,superscriptaddress,nofootinbib]{revtex4}

\usepackage{graphicx}
\usepackage{dcolumn}
\usepackage{bm}
\usepackage{amssymb}
\usepackage{amsfonts}
\usepackage{latexsym}
\usepackage{enumerate}
\usepackage{color} 
\usepackage{amsmath}
\usepackage[dvipdfmx]{epsfig}
\usepackage{times}
\usepackage{algorithm}
\usepackage{algorithmic}
\usepackage{cases}

\newcommand{\red}{\color{black}}
\newtheorem{theorem}{Theorem}

\begin{document}
\widetext
\title{Passive verification protocol for thermal graph states}
\author{Kazuki Akimoto}
\affiliation{Department of Physics, Chuo University, 1-13-27 Kasuga, Bunkyo-ku, Tokyo 112-8551, Japan}
\author{Shunji Tsuchiya}
\email{tsuchiya@phys.chuo-u.ac.jp}
\affiliation{Department of Physics, Chuo University, 1-13-27 Kasuga, Bunkyo-ku, Tokyo 112-8551, Japan}
\author{Ryosuke Yoshii}
\affiliation{Center for Liberal Arts and Sciences, Sanyo-Onoda City University, 1-1-1 Daigaku-Dori, Sanyo-Onoda, Yamaguchi 756-0884, Japan}
\author{Yuki Takeuchi}
\email{yuki.takeuchi.yt@hco.ntt.co.jp}
\affiliation{NTT Communication Science Laboratories, NTT Corporation, 3-1 Morinosato Wakamiya, Atsugi, Kanagawa 243-0198, Japan}

\begin{abstract}
Graph states are entangled resource states for universal measurement-based quantum computation.
Although matter qubits such as superconducting circuits and trapped ions are promising candidates to generate graph states, it is technologically hard to entangle a large number of them due to several types of noise.
Since they must be sufficiently cooled to maintain their quantum properties, thermal noise is one of major ones.
In this paper, we show that for any temperature $T$, the fidelity $\langle G|\rho_T|G\rangle$ between an ideal graph state $|G\rangle$ at zero temperature and a thermal graph state $\rho_T$, which is a graph state at temperature $T$,  can be efficiently estimated by using only one measurement setting.
A remarkable property of our protocol is that it is passive, while existing protocols are active, namely they switch between at least two measurement settings.
Since thermal noise is equivalent to an independent phase-flip error, our estimation protocol also works for that error.
By generalizing our protocol to hypergraph states, we apply our protocol to the quantum-computational-supremacy demonstration with instantaneous quantum polynomial time circuits.
Our results should make the characterization of entangled matter qubits extremely feasible under thermal noise.
\end{abstract}
\maketitle

\section{Introduction}
Quantum computation is expected to outperform classical counterparts~\cite{M16}.
Driven by this expectation, several models for achieving it have been proposed such as the quantum circuit model~\cite{NC00}, adiabatic quantum computation~\cite{FGGS00}, measurement-based quantum computation (MBQC)~\cite{RB01}, and topological quantum computation~\cite{K03}.
MBQC is one of promising universal quantum computing models due to its wide application range, from secure cloud quantum computing~\cite{BFK09,MF13} to quantum error correction~\cite{RHG06}.
In this model, the quantum computation proceeds via adaptive single-qubit measurements on an entangled resource state (e.g., graph states (see Sec.~\ref{II} for its definition)).
Here, ``adaptive" means that each measurement basis can depend on all previous measurement outcomes.

So far, several physical systems have been proposed as candidates for realizing MBQC such as trapped ions~\cite{LJZHMDBBR13} and superconducting qubits~\cite{ABSRLRS18}.
However, at least with current technology, it is hard to generate a large-scale graph state by using them.
An obstacle to its generation is the necessity of cooling.
To maintain their quantum properties, such as superposition and entanglement, they must be cooled below a few tens of millikelvins.
If the cooling is not sufficient, the graph state is affected by thermal noise and consequently becomes a thermal graph state (see Sec.~\ref{II} for the definition).
Several effects of thermal noise on graph states are explored in Ref.~\cite{JDBBR09}.
In particular, a critical temperature at which the graph state becomes a quantum state with only classical correlations is discussed in Ref.~\cite{HV10}.

As the size of MBQC increases, it becomes more important to efficiently check whether a target graph state $|G\rangle$ is precisely prepared, namely to efficiently estimate the fidelity between $|G\rangle$ and an actually prepared state $\rho\equiv\mathcal{E}(|G\rangle\langle G|)$ that suffers from any noise $\mathcal{E}$.
This task is called the verification of graph states.
Although the fidelity can be estimated by using quantum state tomography~\cite{SBRF93,H97,BDPS99} or direct fidelity estimation~\cite{FL11}, these protocols require an exponential number of copies of $\rho$.
Therefore, several efficient verification protocols tailored for graph states have been proposed~\cite{HM15,MNS16,FH17,TM18,TMMMF19,ZH19,ZH19A,MK20}.
In brief, these protocols can be formalized as follows: first, a quantum computer generates some copies of $\rho$ and sends each qubit one by one to a verifier.
Then the verifier randomly chooses a measurement basis and measures the received state $\rho$ in the basis.
He/she repeats the same procedure for all copies of $\rho$.
Finally, by processing all measurement outcomes with a classical computer, he/she outputs an estimated value of (or a lower bound on) the fidelity.
Note that in most cases, only (non-adaptive) single-qubit measurements and efficient classical operations are allowed for the verifier because multi-qubit operations are technologically demanding.
In this paper, we also consider the same restriction on the verifier to simplify the requirement for the verifier as much as possible.

The above existing protocols are active, that is the verifier must switch between some kinds of measurement settings.
In many practical scenarios, the switching of measurement settings could be slow, and in some cases, it may be impossible (e.g., see Ref.~\cite{LZLZ21}).
For example, it would be somewhat demanding to change measurement bases in the IBM quantum cloud service.
It is theoretically and experimentally important to clarify how many measurement settings are required for the verification of graph states.

If any $n$-qubit unitary operator $U$ is allowed for the verifier, the necessary number of measurement settings can be trivially reduced to $1$.
This is because by using the unitary operator $U^\dag$ such that $U|0^n\rangle=|G\rangle$, the verifier can perform the measurement $\{|G\rangle\langle G|,I^{\otimes n}-|G\rangle\langle G|\}$, where $I$ is the two-dimensional identity operator.
However, as we have mentioned above, we would prefer not to allow the verifier to perform such multi-qubit operations.
Furthermore, if the verifier can perform such $U$, then he/she can generate the ideal graph state $|G\rangle$ by his/herself, that is he/she can completely remove the thermal noise, which is unrealistic.
Therefore, we should consider the necessary number of measurement settings under the assumption that the verifier can perform only non-adaptive single-qubit projective measurements.
Unfortunately, under this assumption, at least two measurement settings are required for the verification of any bipartite pure entangled state that includes a subclass of graph states~\cite{LZLZ21}.
Even if adaptive single-qubit projective measurements are allowed for the verifier, at least two measurement settings are still necessary~\cite{ZH19A}.

In this paper, we circumvent the no-go result by assuming that the noise $\mathcal{E}$ is thermal noise.
More precisely, we propose a passive verification protocol for thermal graph states that requires only one measurement setting.
Since the thermal noise is a major obstacle hindering the generation of large-scale graph states with matter qubits, our protocol is still sufficiently practical under this assumption.
Since the thermal noise can be rephrased as an independent phase-flip error, our protocol also works for MBQC with photonic graph states.
Photons are another promising candidate for qubits, and several experiments have been performed to generate photonic graph states~\cite{WRRSWVAZ05,LZGGZYGYP07} and photonic thermal graph states~\cite{AKCACWR14}.
Furthermore, by generalizing our protocol to thermal hypergraph states that are generalizations of thermal graph states (see Sec.~\ref{II} for the definition), we apply our protocol to the demonstration of quantum computational supremacy with instantaneous quantum polynomial time (IQP) circuits~\cite{BMS16}.
The demonstration of quantum computational supremacy is to generate a probability distribution that cannot be efficiently generated with classical computers (under plausible complexity-theoretic assumptions)~\cite{HM17}.
Although its proof-of-principle experiments have already been performed by using random unitary circuits with superconducting qubits~\cite{google,ZC5D2FG4H4L7Q2RS2W4XY7Z8LPZP21} and a boson sampler with squeezed states~\cite{ZWDCPLQWDHHYZLLJGYYWLLLP20}, an experiment with IQP circuits has not yet been performed.
Our protocol should facilitate the realization of large-scale IQP circuits.

The rest of this paper is organized as follows: in Sec.~\ref{II}, we review graph and hypergraph states and their thermal analogues.
In Sec.~\ref{III}, as a main result, we present our passive verification protocol for thermal graph states and show that it is optimal in a sense.
In Sec.~\ref{IV}, we compare our protocol with existing protocols.
In Sec.~\ref{V}, we generalize our protocol to thermal hypergraph states and apply it to the demonstration of quantum computational supremacy with IQP circuits.
Section~\ref{VI} concludes the paper with a brief discussion.
In Appendices A and B, we provide a proof of Theorem~\ref{theorem1}.

\medskip
\section{Thermal graph and hypergraph states}
\label{II}
In this section, we formally define thermal graph and hypergraph states.
To this end, we first review graph~\cite{BR01} and hypergraph states~\cite{RHBM13}.
A graph $G\equiv(V,E)$ is a pair of the set $V$ of $n$ vertices and the set $E$ of edges.
An $n$-qubit graph state $|G\rangle$ corresponding to the graph $G$ is defined as
\begin{eqnarray}
\label{graph}
|G\rangle\equiv\left(\prod_{(i,j)\in E}CZ_{i,j}\right)|+\rangle^{\otimes n},
\end{eqnarray}
where $|+\rangle\equiv(|0\rangle+|1\rangle)/\sqrt{2}$, and $CZ_{i,j}$ is the controlled-$Z$ ($CZ$) gate applied on the $i$th and $j$th qubits.
It is known that Eq.~(\ref{graph}) is a unique common eigenstate with eigenvalue $+1$ of stabilizer operators $\{g_i\}_{i=1}^n$, where
\begin{eqnarray}
\label{gstabilizer}
g_i\equiv X_i\left(\prod_{j:\ (i,j)\in E}Z_j\right)
\end{eqnarray}
for all $1\le i\le n$.
Here, $X_i$ and $Z_j$ are the Pauli-$X$ and $Z$ operators acting on the $i$th and $j$th qubits, respectively, and the product is taken over all vertices $j$ such that $(i,j)\in E$.

Hypergraph states are defined by generalizing graphs to hypergraphs.
A hypergraph $\tilde{G}\equiv(V,\tilde{E})$ is a pair of the set $V$ of $n$ vertices and the set $\tilde{E}$ of hyperedges that can connect more than two vertices, while edges can connect only two vertices.
Let $\tilde{E}$ be the union of $\tilde{E}_2$ and $\tilde{E}_3$, which are sets of hyperedges connecting two and three vertices, respectively~\cite{FN1}.
An $n$-qubit hypergraph state $|\tilde{G}\rangle$ corresponding to the hypergraph $\tilde{G}$ is defined as
\begin{eqnarray}
\label{hyper}
|\tilde{G}\rangle\equiv\left(\prod_{(i,j,k)\in \tilde{E}_3}CCZ_{i,j,k}\right)\left(\prod_{(i,j)\in \tilde{E}_2}CZ_{i,j}\right)|+\rangle^{\otimes n},
\end{eqnarray}
where $CCZ_{i,j,k}$ is the controlled-controlled-$Z$ ($CCZ$) gate applied on the $i$th, $j$th, and $k$th qubits.
From Eq.~(\ref{hyper}), we notice that graph states are special cases of hypergraph states.
When the set $\tilde{E}_3$ is empty, the hypergraph state $|\tilde{G}\rangle$ becomes a graph state.
For the hypergraph state $|\tilde{G}\rangle$, we can define generalized stabilizer operators $\{\tilde{g}_i\}_{i=1}^n$ as follows:
\begin{eqnarray}
\tilde{g}_i\equiv X_i\left(\prod_{j:\ (i,j)\in \tilde{E}_2}Z_j\right)\left(\prod_{(j,k):\ (i,j,k)\in \tilde{E}_3}CZ_{j,k}\right).
\end{eqnarray}
The equality $\tilde{g}_i|\tilde{G}\rangle=|\tilde{G}\rangle$ holds for any $1\le i\le n$, and no other state can satisfy it for all $i$.

Let $\beta\equiv 1/(k_B T)$ with the Boltzmann constant $k_B$ and temperature $T$.
For a Hamiltonian $\mathcal{H}$,
\begin{eqnarray}
\rho_T\equiv\cfrac{e^{-\beta\mathcal{H}}}{\mathcal{Z}},
\end{eqnarray}
where $\mathcal{Z}\equiv{\rm Tr}[e^{-\beta\mathcal{H}}]$ is the partition function, is called the thermal state at temperature $T$.
When 
\begin{eqnarray}
\mathcal{H}=-\sum_{i=1}^ng_i
\end{eqnarray}
and $T\neq 0$, we call $\rho_T$ the thermal graph state, which is a graph state at non-zero temperature.
This is because when the temperature is zero, $\rho_0=|G\rangle\langle G|$.
An experiment has been performed toward the ground-state cooling for the Hamiltonians that are sums of stabilizer operators~\cite{BMSNMCHRZB11}.
In a similar way, we call $\rho_T$ the thermal hypergraph state when 
\begin{eqnarray}
\mathcal{H}=-\sum_{i=1}^n\tilde{g}_i
\end{eqnarray}
and $T\neq 0$.
In this case, $\rho_T$ becomes $|\tilde{G}\rangle\langle\tilde{G}|$ when $T=0$.

These thermal states can also be represented as graph and hypergraph states with an independent phase-flip error (i.e., a Pauli-$Z$ error).
Note that in general, the replacement of the thermal noise with a Pauli-$Z$ error is not possible.
By restring ideal quantum states to graph and hypergraph states, we make the replacement possible.
Let $\mathcal{E}_i^{(p)}(\cdot)\equiv(1-p)(\cdot)+pZ_i(\cdot)Z_i$ be the superoperator realizing the $Z$ error on the $i$th qubit with error probability $p$.
From Ref.~\cite{F15}, when $\mathcal{H}=-\sum_{i=1}^ng_i$,
\begin{eqnarray}
\label{zerror}
\rho_T=\left(\prod_{i=1}^n\mathcal{E}_i^{(p_\beta)}\right)\left(|G\rangle\langle G|\right),
\end{eqnarray}
where $p_\beta\equiv e^{-2\beta}/(1+e^{-2\beta})$.
For the thermal hypergraph states, the same expression holds by replacing $|G\rangle\langle G|$ with $|\tilde{G}\rangle\langle\tilde{G}|$ in Eq.~(\ref{zerror}).
These expressions are useful in evaluating our passive verification protocols.

\begin{figure}[t]
\includegraphics[width=8.5cm,clip]{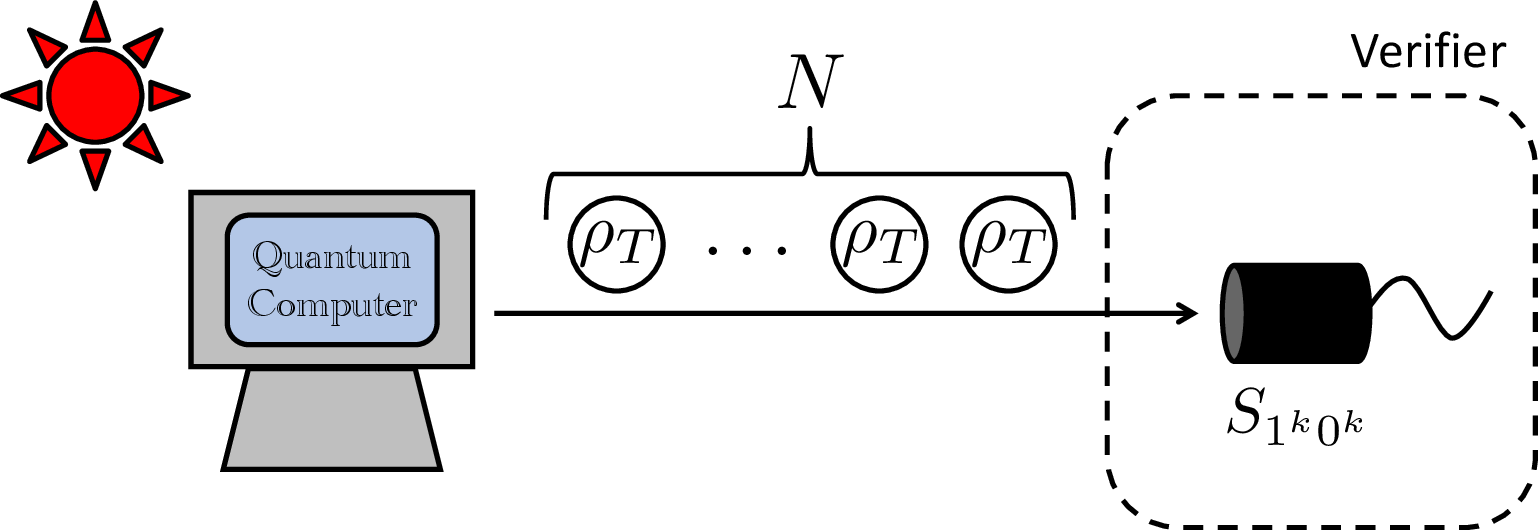}
\caption{Schematic diagram of our passive verification protocol.
A quantum computer affected by thermal noise generates thermal graph states $\rho_T^{\otimes N}$ and sends each qubit one by one.
The verifier just measures $S_{1^k0^k}$ on each received state $\rho_T$ by using only single-qubit Pauli measurements.
No quantum memory is required for the verifier.}
\label{dverification}
\end{figure}

\begin{figure}[t]
\includegraphics[width=6cm,clip]{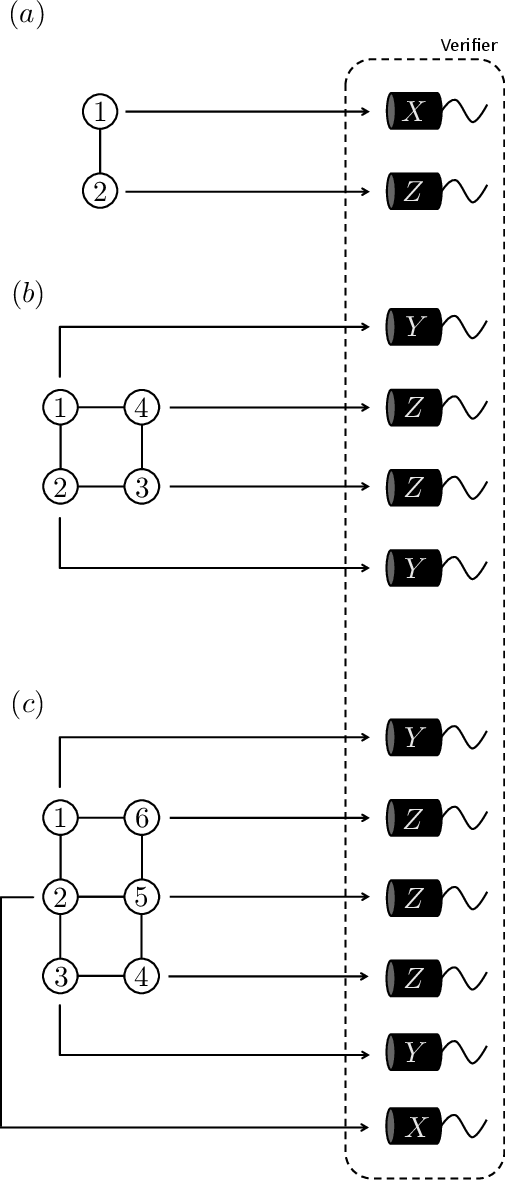}
\caption{Concrete examples of our protocol. Each circle and line represent $|+\rangle$ and the $CZ$ gate, respectively. $X$, $Y$, and $Z$ represent the Pauli-$X$, $Y$, and $Z$ measurements, respectively. The graph states depicted on the left-hand side are the ideal states. (a) $S_{10}=X\otimes Z$. (b) $S_{1100}=Y\otimes Y\otimes Z\otimes Z$. (c) $S_{111000}=-Y\otimes X\otimes Y\otimes Z\otimes Z\otimes Z$. The effect of the minus sign can be incorporated by a classical post-processing.}
\label{CEX}
\end{figure}

\medskip
\section{Passive verification protocol}
\label{III}
In this section, we propose our passive verification protocol for thermal graph states.
Note that since we assume that the thermal graph states are generated due to thermal noise, the temperature $T$ (i.e., the value of $\beta$) is unknown for the verifier.
As a remarkable property, no switching between measurement bases is required for the verifier in our protocol, while previous verification protocols require it, namely they are active.
For simplicity, we assume that the size $n$ of graph states is $2k$ for a natural number $k$.
Let $S_{\bf \ell}\equiv\prod_{i=1}^ng_i^{\ell_i}$ with ${\bf \ell}\equiv \ell_1\ell_2\ldots \ell_n\in\{0,1\}^n$.
Our protocol runs as follows (see also Fig.~\ref{dverification}):
\begin{enumerate}
\item A quantum computer generates $N$ thermal graph states $\rho_T^{\otimes N}$ and sends them to the veirifier.
\item The verifier measures $S_{1^k0^k}$ on each received state $\rho_T$.
Let $o_i\in\{+1,-1\}$ be the $i$th outcome for $1\le i\le N$.
\item The verifier outputs 
\begin{eqnarray}
\label{estimate}
F_{\rm est}\equiv\cfrac{\sum_{i=1}^No_i}{N}
\end{eqnarray}
as an estimated value of the fidelity.
\end{enumerate}
The measurement of $S_{1^k0^k}$ in step 2 can be realized by single-qubit Pauli measurements because $S_{1^k0^k}$ is a tensor product of Pauli operators.
Furthermore, by sequentially sending qubits one by one in step 1, no quantum memory is required for the verifier.
To clarify these properties, we give concrete examples of our protocol in Fig.~\ref{CEX}.

For our protocol, the following theorem holds:
\begin{theorem}
\label{theorem1}
Let $0<\epsilon,\delta<1$, $F_{\rm est}$ be the value defined in Eq.~(\ref{estimate}), and $|G\rangle$ and $\rho_T$ be the $n$-qubit target graph state and the $n$-qubit thermal graph state at unknown temperature $T$, respectively.
When $N=\lceil2/\epsilon^2\log{(2/\delta)}\rceil$ and $n$ is even and at least $4$, the estimated value $F_{\rm est}$ satisfies
\begin{eqnarray}
\label{theorem1eq}
\left|\langle G|\rho_T|G\rangle-F_{\rm est}\right|&\le& \cfrac{n e^{-4\beta}}{2(1+e^{-2\beta})^n}+\epsilon\\
\label{theorem1eq2}
&\le&\cfrac{2}{n}+\epsilon
\end{eqnarray}
with probability at least $1-\delta$.
Here, $\lceil\cdot\rceil$ is the ceiling function.
Particularly, in the limit of large $N$, the inequality $\langle G|\rho_T|G\rangle\ge F_{\rm est}$ holds with unit probability.
\end{theorem}
The proof is given in Appendix A.
Theorem~\ref{theorem1} implies that the more the number $n$ of qubits increases, the more the precision of our protocol improves.
In other words, our protocol is efficient for any temperature $T$.
Although Theorem~\ref{theorem1} is shown for only even $n$, our protocol can be applied when $n$ is odd by simply replacing the ideal state $|G\rangle$ with another graph state $|G\rangle|+\rangle$.
By adding $|+\rangle$, we can always make the number of vertices even.

In step 2, we choose $S_{1^k0^k}$ among $2^n$ kinds of $S_{\bf\ell}$.
This choice is optimal in the sense that it maximally improves the dependence on $T$ of the upper bound in Eq.~(\ref{theorem1eq}).
More formally, we show the following theorem:
\begin{theorem}
\label{theorem2}
Let ${\rm wt}({\bf\ell})\equiv\sum_{i=1}^n\ell_i$ be the Hamming weight for any ${\bf \ell}\in\{0,1\}^n$.
Suppose that we replace $S_{1^k0^k}$ with any $S_{\bf\ell}$ in step 2.
In the limit of large $N$, the upper bound in Eq.~(\ref{theorem1eq}) is
\begin{eqnarray}
\label{gupper}
\cfrac{\left|(n-2{\rm wt}({\bf\ell}))e^{-2\beta}+O(e^{-4\beta})\right|}{(1+e^{-2\beta})^n}
\end{eqnarray}
with unit probability for any fixed natural number $n$.
\end{theorem}
({\it Proof})
In the large limit of $N$, the values $\epsilon$ and $\delta$ become zero, i.e., the estimated value $F_{\rm est}$ becomes ${\rm Tr}[\rho_TS_{\bf\ell}]$.
Therefore, from Eqs.~(\ref{fidelity}) and (\ref{general}) in Appendix A, we immediately obtain Eq.~(\ref{gupper}).
\hspace{\fill}$\blacksquare$

Since Eq.~(\ref{gupper}) is asymptotically minimized when the term of $e^{-2\beta}$ vanishes, Theorem~\ref{theorem2} implies that our protocol is optimized when ${\rm wt}({\bf\ell})=n/2=k$.
Thus $S_{1^k0^k}$ is one of optimal choices.

\begin{figure*}[t]
\includegraphics[width=15cm,clip]{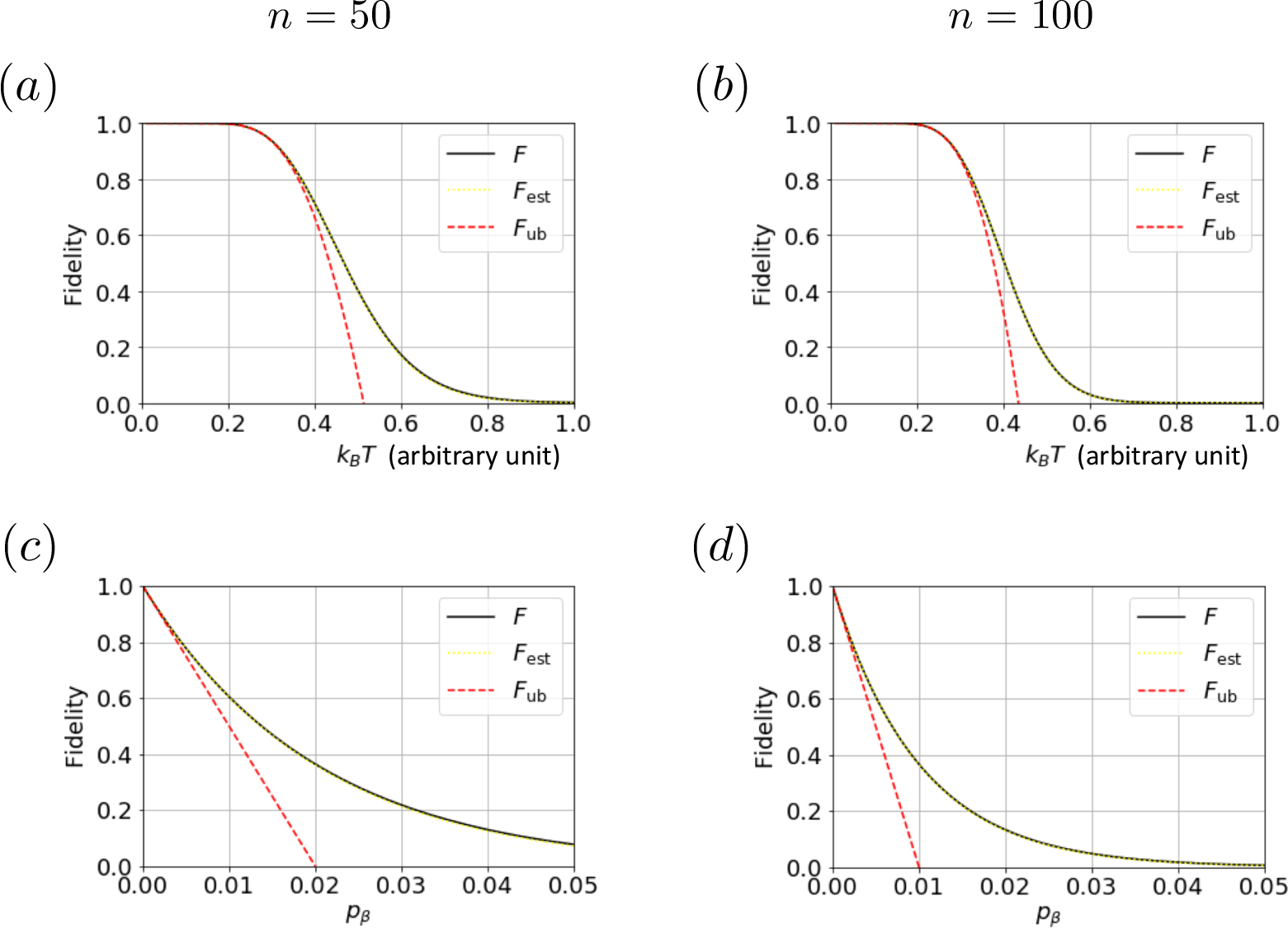}
\caption{Temperature dependence of our estimated value $F_{\rm est}$ and the value $F_{\rm ub}$ obtained with the union bound for $n=50$ and $100$. The {\red solid black, dotted yellow, and dashed red} lines represent the fidelity $F\equiv\langle G|\rho_T|G\rangle$, $F_{\rm est}$, and $F_{\rm ub}$, respectively. In (a) and (b), the label of the horizontal axis is $K_BT$. In (c) and (d), it is converted to $p_\beta$. In (a) and (c) [(b) and (d)], fidelity is plotted in the case of $n=50$ $(100)$.}
\label{comparison}
\end{figure*}

\medskip
\section{Comparison with previous protocols}
\label{IV}
There exist several verification protocols for graph states.
For example, the protocol in Ref.~\cite{MNS16} works for any type of error even if it is a time-correlated one.
However, it requires all of $S_{\bf\ell}$, that is it needs $2^n$ kinds of measurement settings~\cite{FN2}.
Although the necessary number of measurement settings can be improved to $n$ by using the union bound as shown in Ref.~\cite{TM18}, the obtained lower bound $F_{\rm ub}$ on the fidelity becomes loose.
When the protocol is applied to the case of thermal noise with unknown temperature $T$, the lower bound is
\begin{eqnarray}
F_{\rm ub}&=&1-\sum_{i=1}^n\left(1-{\rm Tr}\left[\rho_T\left(\cfrac{I^{\otimes n}+g_i}{2}\right)\right]\right)\\
&=&1-np_{\beta}
\end{eqnarray}
in the limit of large $N$, where we have used Eq.~(\ref{general}) to derive the second equality.
Therefore, by increasing $n$, the lower bound $F_{\rm ub}$ becomes loose.
In contrast, in the limit of large $N$, our lower bound $F_{\rm est}$ becomes tight by increasing $n$ [see Eq.~(\ref{theorem1eq2})].
Fig.~\ref{comparison} gives a comparison between $F_{\rm est}$ and $F_{\rm ub}$ in the limit of large $N$.
It shows that our estimated value $F_{\rm est}$ is close to the actual value $F\equiv\langle G|\rho_T|G\rangle$ even when the number $n$ of qubits and $k_BT$ (i.e., $p_{\beta}$) are large.
On the other hand, $F_{\rm ub}$ becomes distinct from $F$ when $k_BT$ is large (i.e., $p_{\beta}$ is large).

The number of required measurement settings was further improved by restricting the target states.
In particular, only two measurement settings are sufficient when the target state is any bipartite graph state~\cite{HM15}.
Recently, Li {\it et al.} generalized this result to any bipartite pure entangled state and showed that two settings cannot be improved to one~\cite{LZLZ21}.
They also showed that any bipartite pure product state can be verified with a single measurement setting if local projective measurements are allowed for the verifier~\cite{LZLZ21}.
Our result implies that a single measurement setting is sufficient even for graph states if the type of noise is restricted.
By assuming thermal noise (i.e., the independent $Z$ error), we circumvent the no-go result in Ref.~\cite{LZLZ21}.

\medskip
\section{Generalizations to thermal hypergraph states}
\label{V}
\begin{figure}[t]
\includegraphics[width=8.5cm,clip]{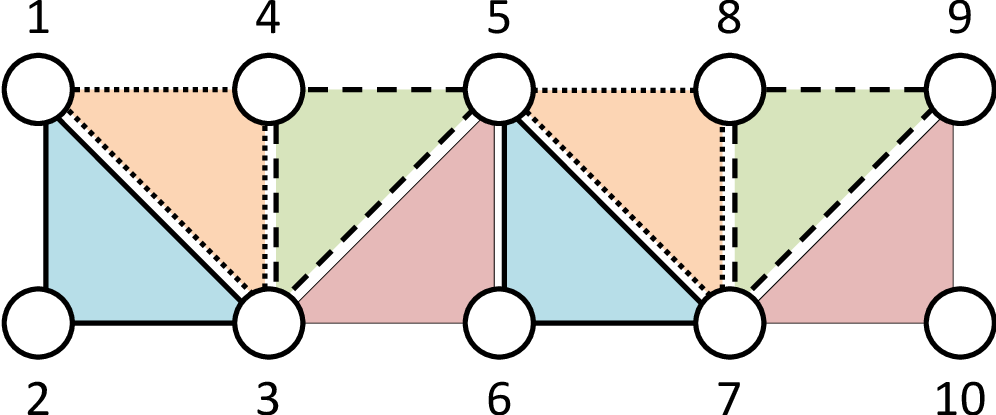}
\caption{A $10$-qubit hypergraph state with $\tilde{E}_2=\emptyset$. For $1\le i\le 10$, the $i$th circle represents the $i$th qubit. The solid-line blue, dotted-line orange, dashed-line green, and thin-solid-line red triangles represent $CCZ$ gates applied on $\{(4j-3,4j-2,4j-1)\}_{j=1}^2$, $\{(4j-3,4j-1,4j)\}_{j=1}^2$, $\{(4j-1,4j,4j+1)\}_{j=1}^2$, and $\{(4j-1,4j+1,4j+2)\}_{j=1}^2$, respectively.}
\label{hyper10}
\end{figure}

In this section, we generalize our protocol to a restricted class of $n$-qubit thermal hypergraph states that become, at the temperature $T=0$, the hypergraph states in Eq.~(\ref{hyper}) such that $\tilde{E}_3$ is the union of $\{(4j-3,4j-2,4j-1)\}_{j=1}^{\lceil(n+1)/4\rceil}$, $\{(4j-3,4j-1,4j)\}_{j=1}^{\lceil n/4\rceil}$, $\{(4j-1,4j,4j+1)\}_{j=1}^{\lceil(n-1)/4\rceil}$, and $\{(4j-1,4j+1,4j+2)\}_{j=1}^{\lceil(n-2)/4\rceil}$.
Note that $\tilde{E}_2$ is arbitrary, and we assume that $n$ is even.
For clarity, we depict such a hypergraph state with $n=10$ and $\tilde{E}_2=\emptyset$ in Fig.~\ref{hyper10}.

Let $\tilde{S}_{\bf\ell}\equiv\prod_{i=1}^n\tilde{g}_i^{\ell_i}$ with ${\bf\ell}=\ell_1\ell_2\ldots\ell_n\in\{0,1\}^n$.
In the case of hypergraph states mentioned above, by setting ${\bf\ell}=0101\ldots01=(01)^{n/2}$, we can obtain
\begin{eqnarray}
\tilde{S}_{(01)^{n/2}}=\prod_{(i,j)\in\tilde{E}_2}CZ_{i,j}\left(\prod_{k=1}^{n/2}X_{2k}\right)CZ_{i,j}.
\end{eqnarray}
When $i=2k$ $(j=2k)$, the equality $CZ_{i,j}X_{2k}CZ_{i,j}=X_iZ_j$ $(Z_iX_j)$ holds.
Otherwise, $CZ_{i,j}X_{2k}CZ_{i,j}=X_{2k}$.
Therefore, $\tilde{S}_{(01)^{n/2}}$ is a tensor product of $n$ Pauli operators (up to $\pm 1$), and just half of them are the Pauli-$X$ or $Y$.
As a result, from the argument in Appendix A, by replacing $S_{1^k0^k}$ with $\tilde{S}_{(01)^{n/2}}$ in step 2 of our protocol, we can also show that Theorem~\ref{theorem1} holds for the thermal hypergraph states.

\begin{figure}[t]
\includegraphics[width=8.5cm,clip]{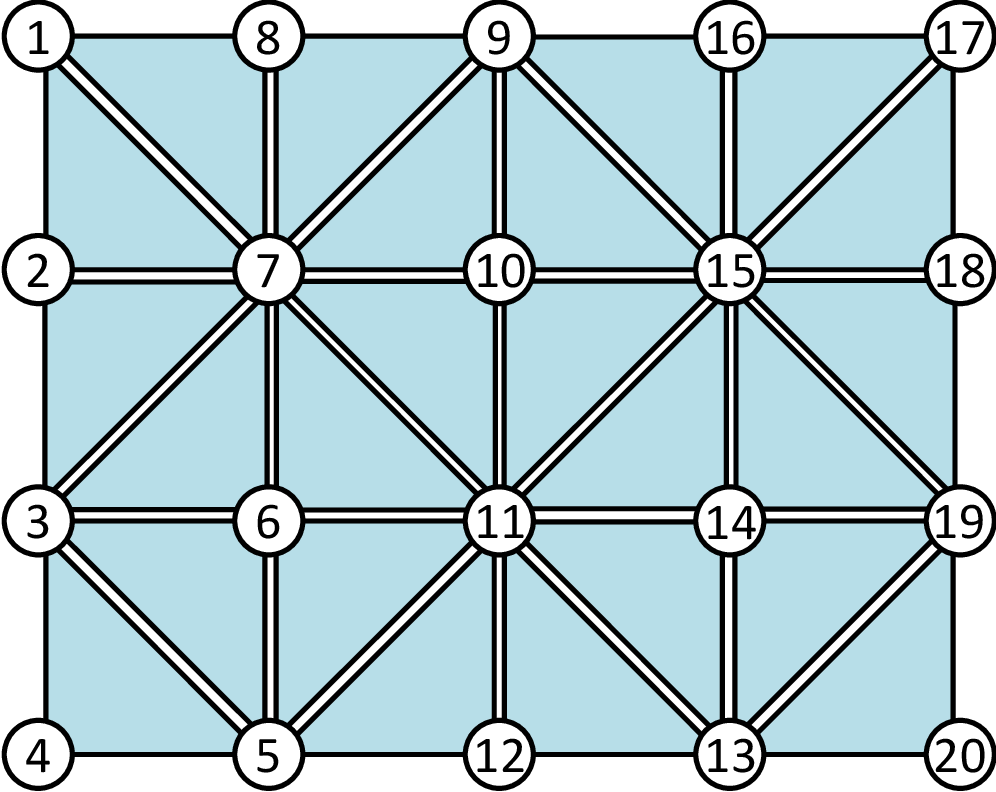}
\caption{A $20$-qubit hypergraph state with four column qubits. For $1\le i\le 20$, the $i$th circle represents the $i$th qubit. The solid-line blue triangles represent $CCZ$ gates.}
\label{hyper20}
\end{figure}

Hypergraph states in the above class have only two qubits in the column direction while they can have any number of qubits in the row direction (see Fig.~\ref{hyper10}).
The above argument can also be applied even if we increase the number of qubits in the column direction to more than two.
As an example, let us consider the $20$-qubit hypergraph state in Fig.~\ref{hyper20}.
Since $\tilde{S}_{(01)^{10}}=\prod_{i=1}^{10}X_{2i}$, we can show that Theorem~\ref{theorem1} holds for this hypergraph state.
Therefore, our verification protocol is applicable to hypergraph states from which we can obtain the Union Jack states~\cite{MM16}, which are universal resource states for MBQC, under postselection.

Our verification protocol can be applied to the demonstration of quantum computational supremacy with IQP circuits~\cite{BMS16} under conjectures.
An $n$-qubit IQP circuit is a quantum circuit such that the initial state and measurements are $|+\rangle^{\otimes n}$ and $X$-basis measurements, respectively, and the applied quantum gate $D$ is any diagonal unitary in the $Z$ basis~\cite{SB09}.
For IQP circuits with $D$ consisting of $Z$, $CZ$, and $CCZ$ gates, Bremner, Montanaro, and Shepherd showed that it is hard to efficiently simulate classically any probability distribution $\{q_z\}_{z\in\{0,1\}^n}$ obtained from IQP circuits~\cite{BMS16}.
More precisely, it is hard to generate $\{p_z\}_{z\in\{0,1\}^n}$ in classical polynomial time such that $\sum_{z\in\{0,1\}^n}|q_z-p_z|\le1/192$.
Therefore, the generation of a probability distribution $\{q'_z\}_{z\in\{0,1\}^n}$ such that $\sum_{z\in\{0,1\}^n}|q_z-q'_z|\le1/192$ is called the demonstration of quantum computational supremacy.
Since IQP circuits just measure hypergraph states in the $X$ basis when $D$ consists of $Z$, $CZ$, and $CCZ$ gates, their result shows that if appropriate hypergraph states can be prepared with sufficiently high fidelity, then quantum computational supremacy is successfully demonstrated.

Their hardness proof is based on the anticoncentration lemma and two plausible complexity-theoretic conjectures (i.e., the average-case ${\sf \#P}$-hardness of the approximation of output probabilities $q_{0^n}$ for random $D$'s and infiniteness of the polynomial hierarchy).
The lemma implies that for any $z$, when $D$ is randomly chosen, the probability of $q_z$ being larger than a certain value is at least a constant.
It holds even if qubits on which $CCZ$ gates are applied are fixed as long as qubits on which $Z$ and $CZ$ gates are applied are randomly chosen.
In other words, hypergraph states with a fixed $\tilde{E}_3$ are sufficient to show the anticoncetration lemma.
This is why our hypergraph states mentioned above can be used to demonstrate quantum computational supremacy under the assumption that the two conjectures hold for our class of hypergraph states with the fixed $\tilde{E}_3$.
A similar assumption was used in Ref.~\cite{MTH17}.

Although several certification protocols to check whether the demonstration of quantum computational supremacy is correctly achieved have been proposed for IQP circuits~\cite{HKSE17,MTH17,TM18,KD19}, they are active; that is, some switching between quantum operations is required for a certifier.
In contrast, we propose a passive certification protocol under the assumption that the noise is thermal noise or an independent phase-flip error.
For simplicity, we set $n\ge 4\times 10^5$, $\epsilon=10^{-6}$, and $\delta=10^{-2}$ in Theorem~\ref{theorem1}.
Our certified quantum computational supremacy protocol proceeds as follows:
\begin{enumerate}
\item An experimentalist runs a quantum computer to generate an $n$-qubit thermal hypergraph state $\rho_T$ and measures $\tilde{S}_{(01)^{n/2}}$.
He/she repeats the same procedure $N(\le 1.06\times 10^{13})$ times.
Let $o_i\in\{+1,-1\}$ be the $i$th outcome for $1\le i\le N$.
\item The experimentalist calculates
\begin{eqnarray}
F_{\rm est}=\cfrac{\sum_{i=1}^No_i}{N}.
\end{eqnarray}
\item If $F_{\rm est}-2/n\ge0.999995$, the experimentalist generates $\{q'_z\}_{z\in\{0,1\}^n}$ by generating $\rho_T$'s again and then measuring them in the $X$ basis.
Otherwise, he/she declares that the precision of the quantum computer is not enough to demonstrate quantum computational supremacy.
\end{enumerate}

We show that when $F_{\rm est}-2/n\ge0.999995$, the generated probability distribution $\{q'_z\}_{z\in\{0,1\}^n}$ satisfies $\sum_{z\in\{0,1\}^n}|q_z-q'_z|\le1/192$ with probability of $0.99$, that is the experimentalist succeeds in demonstrating quantum computation supremacy.
Note that if the temperature is $0$, the inequality $F_{\rm est}-2/n\ge0.999995$ can be satisfied with unit probability because $F_{\rm est}=1$ and $n\ge 4\times 10^5$.
From Ref.~\cite{NC00} and the Fuchs-van de Graaf inequality~\cite{FG99},
\begin{eqnarray}
\sum_{z\in\{0,1\}^n}|q_z-q'_z|&\le&\left|\left||\tilde{G}\rangle\langle \tilde{G}|-\rho_T\right|\right|\\
&\le&2\sqrt{1-\langle\tilde{G}|\rho_T|\tilde{G}\rangle},
\end{eqnarray}
where $||\cdot||$ is the trace norm.
Therefore, from Theorem~\ref{theorem1} with $\epsilon=10^{-6}$ and $\delta=10^{-2}$, the required number $N$ of samples is at most $1.06\times 10^{13}$, and
\begin{eqnarray}
\label{supremacy}
\sum_{z\in\{0,1\}^n}|q_z-q'_z|\le2\sqrt{1+10^{-6}-\left(F_{\rm est}-\cfrac{2}{n}\right)}
\end{eqnarray}
with probability at least $0.99$.
To satisfy that the right-hand side of Eq.~(\ref{supremacy}) is at most $1/192$, it is sufficient to satisfy
\begin{eqnarray}
F_{\rm est}-\cfrac{2}{n}&\ge&1+10^{-6}-\cfrac{1}{384^2}\\
&=&0.999994\ldots
\end{eqnarray}

\medskip
\section{Conclusion and discussion}
\label{VI}
We have proposed a passive verification protocol for thermal graph and hypergraph states.
As a remarkable property, our verification protocol requires only one measurement setting.
This passiveness cannot be obtained when the noise is general, even if any single-qubit projective measurements are allowed for the verifier~\cite{ZH19A,LZLZ21}.
We circumvent their no-go result by assuming thermal noise (or independent phase-flip error), which is a major obstacle to the generation of large-scale entangled states.
Note that for Fock-basis photonic quantum states, a passive verification protocol has already been proposed~\cite{CGKM21}.
Particularly, under the two conjectures, our verification protocol for thermal hypergraph states can be used to demonstrate quantum computational supremacy in a certifiable manner.
As another application, our verification protocol can also be used as a quantum sensing protocol to estimate unknown temperature $T$.
This is because the fidelity $\langle G|\rho_T|G\rangle$ is uniquely determined by $n$ and $\beta$ [see Eq.~(\ref{fidelity}) and Figs.~\ref{comparison} (a) and (b)].
Note that in this application, the product state $|G\rangle=|+\rangle^{\otimes n}$ is sufficient as an input.

It would be interesting to consider whether our protocol can be generalized to other classes of quantum states such as weighted graph states~\cite{HDERNB05,HCDB07} and/or other types of errors such as depolarizing noises and noises induced by the finite lifetime of matter qubits.
Recently, Ref.~\cite{YS22} proposed a one-shot protocol to decide whether an error rate is lower than a constant value $a$ or larger than another constant value $b$ for any bounded-degree periodic graph states with depolarizing noise.
Their idea may be useful in generalizing our passive verification protocol, but we leave that as a future work.

For simplicity, we have assumed that the verifier's measurements are perfect.
However, independent phase-flip errors on the measurement apparatuses can be allowed if the error probabilities are the same for all apparatuses.
Since measurement errors affect the performance of our verification protocol, it should be important to also take other measurement errors into account.
Although an {\it active} error-tolerant verification protocol has already been proposed~\cite{FH17}, it is still open whether it can be combined with our protocol to construct a {\it passive} error-tolerant one.

\medskip
\section*{ACKNOWLEDGMENTS}
We thank Shion Yamashika for fruitful discussions about Appendix A.
ST is supported by the Japan Society for the Promotion of Science Grant-in-Aid for Scientific Research (KAKENHI Grant No.~19K03691).
RY is supported by JSPS Grant-in-Aid for Scientific Research (KAKENHI Grant No.~19K14616 and 20H01838).
YT is supported by MEXT Quantum Leap Flagship Program (MEXT Q-LEAP) Grant Number JPMXS0118067394 and JPMXS0120319794 and JST [Moonshot R\&D -- MILLENNIA Program] Grant Number JPMJMS2061.

\medskip
\section*{APPENDIX A: Proof of Theorem~\ref{theorem1}}
\label{A}
In this appendix, we prove Theorem~\ref{theorem1}.
First, we derive a value to which the estimated value $F_{\rm est}$ in Eq.~(\ref{estimate}) converges.
Let $m_+$ and $m_-(=N-m_+)$ be the numbers of $+1$ and $-1$ outcomes, respectively.
From Eq.~(\ref{estimate}),
\begin{eqnarray}
F_{\rm est}=(+1)\cdot\cfrac{m_+}{N}+(-1)\cdot\cfrac{m_-}{N}.
\end{eqnarray}
Therefore, it converges to ${\rm Tr}[\rho_TS_{1^k0^k}]$, and thus the Hoeffding inequality~\cite{H63} guarantees that when $N=\lceil2/\epsilon^2\log{(2/\delta)}\rceil$, the inequality
\begin{eqnarray}
\label{Hoeffding}
\left|{\rm Tr}[\rho_TS_{1^k0^k}]-F_{\rm est}\right|\le\epsilon
\end{eqnarray}
holds with probability at least $1-\delta$.
By using the triangle inequality and Eq.~(\ref{Hoeffding}),
\begin{eqnarray}
\nonumber
&&\left|\langle G|\rho_T|G\rangle-F_{\rm est}\right|\\
\nonumber
&\le&\left|\langle G|\rho_T|G\rangle-{\rm Tr}[\rho_TS_{1^k0^k}]\right|+\left|{\rm Tr}[\rho_TS_{1^k0^k}]-F_{\rm est}\right|\\
\\
\label{upper1}
&\le&\left|\langle G|\rho_T|G\rangle-{\rm Tr}[\rho_TS_{1^k0^k}]\right|+\epsilon
\end{eqnarray}
with probability at least $1-\delta$.

The remaining task is to upper bound $\left|\langle G|\rho_T|G\rangle-{\rm Tr}[\rho_TS_{1^k0^k}]\right|$.
To this end, we calculate $\langle G|\rho_T|G\rangle$ and ${\rm Tr}[\rho_TS_{1^k0^k}]$ in order.
By substituting Eq.~(\ref{zerror}) for $\rho_T$, we obtain
\begin{eqnarray}
\label{fidelity}
\langle G|\rho_T|G\rangle=(1-p_\beta)^n=1/(1+e^{-2\beta})^n.
\end{eqnarray}
This is because if at least one $Z$ error occurs on the graph state $|G\rangle$, it becomes orthogonal to the ideal state $|G\rangle$.

To calculate ${\rm Tr}[\rho_TS_{1^k0^k}]$, we derive a general expression for ${\rm Tr}[\rho_TS_{\bf \ell}]$ with any ${\bf \ell}\in\{0,1\}^n$.
Recall that $S_{\bf \ell}$ is a tensor product of $n$ Pauli operators, i.e., $S_{\bf \ell}=(-1)^s\otimes_{i=1}^n\sigma_i$ with  $\sigma_i\in\{X,Y,Z,I\}$ and $s\in\{0,1\}$.
Here, $Y$ is the Pauli-$Y$ operator.
Let us call $\sigma_i$ the $i$th operator in $S_{\bf\ell}$.
From Eq.~(\ref{gstabilizer}), we notice that the $i$th operator in $S_{\bf \ell}$ is $X$ or $Y$ if and only if $\ell_i=1$.
In other words, when $\ell_i=0$, the $i$th operator is $Z$ or $I$.
The thermal noise can be considered as an independent phase-flip error as shown in Eq.~(\ref{zerror}), and the phase-flip error is detected by $X$ and $Y$ measurements.
Since the $m$ phase-flip errors occur with probability $(1-p_\beta)^{n-m}p_\beta^m$, using Eq.~(\ref{zerror}), we obtain a general expression for ${\rm Tr}[\rho_T S_{\bf \ell}]$ as
\begin{align}
{\rm Tr}[\rho_T S_{\bf \ell}]=\sum_{m=0}^n C(m) (1-p_\beta)^{n-m}p_{\beta}^m,\label{general}\\
C(m)\equiv\sum_{j=w({\bf\ell},m,n)}^{f({\bf\ell},m)}(-1)^j\binom{{\rm wt}(\bf\ell)}{j}\binom{n-{\rm wt}(\bf\ell)}{m-j},\label{generalC}
\end{align}
where ${\rm wt}(\ell)\equiv\sum_{i=1}^n\ell_i$ is the Hamming weight,
$f({\bf \ell},m)\equiv{\rm min}\{{\rm wt}({\bf\ell}),m\}$, and $w({\bf \ell},m,n)\equiv{\rm max}\{0,{\rm wt}({\bf\ell})+m-n\}$. We assumed $\binom{0}{0}=1$.
$C(m)$ is the summation of the outcomes of the measurements of $S_{\bf\ell}$ for the cases of $m$ phase-flip errors. The term $\binom{{\rm wt}(\bf\ell)}{j}\binom{n-{\rm wt}(\bf\ell)}{m-j}$ is the number of patterns where $j$ phase-flip errors among $m$ phase-flip errors occur on the ${\rm wt}({\bf\ell})$ qubits such that $\ell_i=1$, and the other $m-j$ phase-flip errors occur on the other $(n-{\rm wt}({\bf\ell}))$ qubits such that $\ell_i=0$. The factor $(-1)^{j}$ corresponds to the outcome of the measurement of $S_{\bf\ell}$ that is $+1$ $(-1)$ if the number $j$ of errors among the ${\rm wt}(\bf\ell)$ qubits is even (odd).
To make our argument clearer, we give the concrete derivation of Eq.~(\ref{general}) for the ideal state $CZ|++\rangle$ in Appendix B.

$C(0)=1$ represents that the outcome of the measurement of $S_{\bf\ell}$ is always $+1$ when no error occurs.
This is because $|G\rangle$ is stabilized by any $S_{\bf\ell}$.
Note that since $m=0$ does not imply $T=0$, it does not mean the ideal case.
We here just claim that when the error probability is $p_\beta$, no error occurs on all $n$ qubits with probability $(1-p_\beta)^n$.

We find that $C(m)$ is equivalent to the coefficient of the term $x^{n-m}y^m$ in the expansion of the polynomial $(x-y)^{{\rm wt}({\bf\ell})}(x+y)^{n-{\rm wt}({\bf\ell})}$. Therefore, substituting $x=1-p_{\beta}$ and $y=p_{\beta}$ in $(x-y)^{{\rm wt}({\bf\ell})}(x+y)^{n-{\rm wt}({\bf\ell})}$, we finally obtain
\begin{eqnarray}
\nonumber
&&{\rm Tr}[\rho_T S_{\bf \ell}]\\
&=&[(1-p_{\beta})-p_{\beta}]^{{\rm wt}({\bf\ell})}[(1-p_{\beta})+p_{\beta}]^{n-{\rm wt}({\bf\ell})}\\
&=&(1-2p_{\beta})^{{\rm wt}({\bf\ell})}.
\label{generalshort}
\end{eqnarray}
At zero temperature, i.e., the ideal case $\rho_{0}=|G\rangle\langle G|$, using $p_\beta=0$, we obtain ${\rm Tr}[\rho_T S_{\bf \ell}]=1$ from Eq.~(\ref{generalshort}) as it should be.

We confirm that Eqs.~(\ref{general}), (\ref{generalC}), and (\ref{generalshort}) include the cases of ${\rm wt}({\bf\ell})=n$ and ${\rm wt}({\bf\ell})=0$. Setting ${\rm wt}({\bf\ell})=n$ in them, we obtain
\begin{eqnarray}
{\rm Tr}\left[\rho_T S_{\bf \ell}\right]&=&\sum_{m=0}^n(-1)^m\binom{n}{m}(1-p_{\beta})^{n-m}p_{\beta}^m\\
&=&[(1-p_{\beta})-p_{\beta}]^n=(1-2p_{\beta})^n,
\label{generaln}
\end{eqnarray}
where we have used the binomial theorem. This is the expected result. 
Analogously, setting ${\rm wt}({\bf\ell})=0$ in them, we obtain
\begin{eqnarray}
{\rm Tr}\left[\rho_T S_{\bf \ell}\right]&=&\sum_{m=0}^n\binom{n}{m}(1-p_{\beta})^{n-m}p_{\beta}^m\\
&=&[(1-p_{\beta})+p_{\beta}]^n=1.
\label{general0}
\end{eqnarray}
This is consistent with the fact that the outcome of the measurement of $S_{0^n}=I^{\otimes n}$ is always unity.

From Eq.~(\ref{generalshort}), when $n=2k$ and ${\bf \ell}=1^k0^k$, 
\begin{eqnarray}
{\rm Tr}[\rho_TS_{1^k0^k}]=[(1-p_{\beta})^2-p_{\beta}^2]^k=\cfrac{(1-e^{-4\beta})^k}{(1+e^{-2\beta})^n}.
\label{expectation}
\end{eqnarray}
If one expands the middle expression in Eq.~(\ref{expectation}), one finds that $C(m)=0$ when $m$ is odd.
By using Eqs.~(\ref{fidelity}) and (\ref{expectation}),
\begin{eqnarray}
\left|\langle G|\rho_T|G\rangle-{\rm Tr}[\rho_TS_{1^k0^k}]\right|&=&\cfrac{1-(1-e^{-4\beta})^k}{(1+e^{-2\beta})^n}\\
&\le&\cfrac{1-(1-ke^{-4\beta})}{(1+e^{-2\beta})^n}\\
\label{upper2}
&=&\cfrac{ne^{-4\beta}}{2(1+e^{-2\beta})^n}\\
\label{upper3}
&\le&\cfrac{2}{n}.
\end{eqnarray}
Finally, we combine Eqs.~(\ref{upper1}), (\ref{upper2}), and (\ref{upper3}) and obtain Eqs.~(\ref{theorem1eq}) and (\ref{theorem1eq2}).

In the limit of large $N$, the values $\epsilon$ and $\delta$ become zero, i.e., $F_{\rm est}={\rm Tr}[\rho_TS_{1^k0^k}]$.
Therefore, from Eqs.~(\ref{fidelity}) and (\ref{expectation}),
\begin{eqnarray}
\langle G|\rho_T|G\rangle\ge{\rm Tr}[\rho_TS_{1^k0^k}]=F_{\rm est}.
\end{eqnarray}

\medskip
\section*{APPENDIX B: Derivation of Eq.~(\ref{general}) for $CZ|++\rangle$}
In this appendix, we derive the value of ${\rm Tr}[\rho_TS_{\bf\ell}]$ when the ideal state is $|G\rangle=CZ|++\rangle$.
From Eq.~(\ref{zerror}), the thermal graph state $\rho_T$ is
\begin{eqnarray}
\nonumber
&&(1-p_\beta)^2|G\rangle\langle G|\\
\nonumber
&&+p_\beta(1-p_\beta)\left(Z_1|G\rangle\langle G|Z_1+Z_2|G\rangle\langle G|Z_2\right)\\
\label{CEX2}
&&+p_\beta^2Z_1Z_2|G\rangle\langle G|Z_1Z_2,
\end{eqnarray}
where $Z_i$ is the Pauli-$Z$ operator on the $i$th qubit for $i\in\{1,2\}$.
Regardless of the value of $\ell$, the ideal state $|G\rangle\langle G|$ satisfies ${\rm Tr}[|G\rangle\langle G|S_\ell]=1$.
Therefore,
\begin{eqnarray}
\label{C1}
{\rm Tr}\left[(1-p_\beta)^2|G\rangle\langle G|S_{\bf\ell}\right]=C(0)(1-p_\beta)^2,
\end{eqnarray}
which corresponds to the term of $m=0$ in Eq.~(\ref{general}).

Next, we consider the second and third terms in Eq.~(\ref{CEX2}), which correspond to terms of $m=1$ and $m=2$ in Eq.~(\ref{general}), respectively.
The values of ${\rm Tr}[Z_1|G\rangle\langle G|Z_1S_{\bf\ell}]$, ${\rm Tr}[Z_2|G\rangle\langle G|Z_2S_{\bf\ell}]$, and ${\rm Tr}[Z_1Z_2|G\rangle\langle G|Z_1Z_2S_{\bf\ell}]$ depend on ${\bf\ell}$.
For example, when ${\rm wt}({\bf\ell})=1$,
\begin{eqnarray}
{\rm Tr}\left[\left(Z_1|G\rangle\langle G|Z_1+Z_2|G\rangle\langle G|Z_2\right)S_{\bf\ell}\right]&=&0\\
{\rm Tr}\left[Z_1Z_2|G\rangle\langle G|Z_1Z_2S_{\bf\ell}\right]&=&-1.
\end{eqnarray}
Therefore,
\begin{eqnarray}
\nonumber
&&{\rm Tr}\left[p_\beta(1-p_\beta)\left(Z_1|G\rangle\langle G|Z_1+Z_2|G\rangle\langle G|Z_2\right)S_{\bf\ell}\right]\\
\label{C2}
&=&\left[\sum_{j=w({\bf\ell},1,2)}^{f({\bf\ell},1)}(-1)^j\binom{1}{j}\binom{1}{1-j}\right](1-p_\beta)p_\beta,
\end{eqnarray}
where $w({\bf\ell},1,2)=0$ and $f({\bf\ell},1)=1$, and
\begin{eqnarray}
\nonumber
&&{\rm Tr}\left[p_\beta^2Z_1Z_2|G\rangle\langle G|Z_1Z_2S_{\bf\ell}\right]\\
\label{C3}
&=&\left[\sum_{j=w({\bf\ell},2,2)}^{f({\bf\ell},2)}(-1)^j\binom{1}{j}\binom{1}{2-j}\right]p_\beta^2,
\end{eqnarray}
where $w({\bf\ell},2,2)=f({\bf\ell},2)=1$.
By combining Eqs.~(\ref{C1}), (\ref{C2}), and (\ref{C3}), we actually obtain Eq.~(\ref{general}).
For the cases of ${\rm wt}({\bf\ell})=0$ and ${\rm wt}({\bf\ell})=2$, a similar argument holds.


\begin{thebibliography}{99}
\bibitem{FN1}In the definition of hypergraphs, hyperedges that connect more than three vertices are also allowed. However, for our purpose, it is unnecessary to consider such hyperedges.
\end{thebibliography}
\end{document}